\documentclass[epsf,oneside,12pt]{article}
\input psfig.sty
\usepackage{amsmath,eufrak}
\include{amsfonts}
\newcommand{\beq}[2]{\begin{equation}#1\label{#2}\end{equation}}
\newcommand{\ceq}[1]{(\ref{#1})}
\newcommand{\mbd}[1]{\mbox{\bf #1}}

\newcommand{\br}{\mbd{r}}

\newfont{\mbld}{cmbx10 scaled 800}
\newfont{\cab}{cmsy10 scaled 1200}
\newfont{\scab}{cmsy10 scaled 1000}
\newfont{\bcall}{cmbsy10 scaled 1200}

\begin{document}
\title{Chern-Simons Field Theories with Non-semisimple Gauge Group of Symmetry}
\author{Franco Ferrari\\
{\it Institute of Physics, University of Szczecin, ul. Wielkopolska 15,}\\
{\it 70-451 Szczecin, Poland}\thanks{e-mail:
fferrari@univ.szczecin.pl}.}


\maketitle
\renewcommand{\baselinestretch}{1.5}

\abstract{
Subject of this work is a class of Chern-Simons field theories with
non-semisimple gauge group, which may well be considered as the most 
straightforward generalization of an Abelian Chern-Simons field theory.
As a matter of fact these theories, which are characterized by a 
non-semisimple group of gauge symmetry,
have
cubic interactions like those of
non-abelian Chern-Simons field theories, but are free from
radiative corrections. Moreover, at the tree level in the perturbative
expansion,
 there
are only two connected tree diagrams, corresponding to the
propagator and to the three vertex originating from the cubic interaction
terms.
For such theories it is derived here a set of BRST invariant
observables, which lead to metric
 independent amplitudes.
The vacuum expectation values of these observables can be computed exactly.
From their expressions it is possible to
isolate the Gauss linking number
and an invariant of the Milnor type, which describes
the topological relations among three or more closed curves.}
\vfill\eject
\pagestyle{plain}
\section{Foreword}
In several situations it has been experimentally observed that the
topological properties of certain physical systems may influence their
behavior to a relevant extent. This is for instance the case of vortex
structures in nematic liquid crystals \cite{nemliq}
and in $^3He$ superfluids \cite{supflu}. Other
examples are provided by polymers \cite{wasserman}
 or by the lowest lying excitations of
two-dimensional electron gases, which have topological non-trivial
configurations at some filling fractions \cite{quahaleff}.
In the investigation of phenomena
related to the presence of topological constraints in physical systems, the
use of
quantum or statistical mechanical models
coupled to abelian Chern--Simons (C-S) field
theories \cite{chesim}
has been particularly successful.
One reason of this success is the fact that abelian models
do not require a complex mathematical
treatment as their non-abelian counterparts and thus
their physical meaning is more transparent.

Motivated by applications in polymer physics \cite{ferannphys,kolvil,kle}
the aim of
this work is the construction
of topological field theories
with non-trivial cubic interactions similar to those of
non-abelian C-S field theories, but which still retain
the
simplicity of the abelian case. 
For this purpose,
suitable candidates are  C-S field theories with non-semisimple
 group of
gauge symmetry. Roughly speaking,   Lie algebras
 associated to
 non-semisimple
groups contain non-trivial Abelian ideals, so that one could
 expect on this ground that at least part of the observables of
 these theories should have 
``Abelian'' characteristics.

Chern-Simons field theories and, more in general, gauge field theories
with non-semisimple groups of symmetry,
have been already proposed in \cite{wit,tse}.
Here it is picked up 
a particular class of such theories with the main property of being free
from radiative corrections. 
Also at the tree level in the perturbative expansion several
simplifications occur and it is possible to show that
there are
only two connected diagrams, the propagators 
and the three vertices corresponding to the fields' self-interactions.
This situation is reminiscent to that of an Abelian field theory, in which
there is just one connected diagram, namely the propagator.
 Most interestingly, the theories discussed in this work admit
a set of observables which resemble Abelian Wilson loops
 and lead to metric independent
amplitudes. With respect to standard Wilson loops, these observables contain
 extra terms,
which are required to enforce  BRST invariance.
From their vacuum expectation values, which are computed
exactly, it is possible to
isolate a topological invariant, which describes the
topological properties of three or more closed loops.

The material presented in this paper is divided as follows.
In Section 2 a class of C-S field theories with non-semisimple group of
symmetry is introduced, which consists in a set of abelian
BF--models \cite{bfmod} coupled together by
cubic interaction terms. The BRST quantization of these theories 
is discussed using the covariant gauge of Lorentz in order to fix gauge
invariance.
Further, it is shown that radiative corrections are absent and
 that there are only two connected Feynman diagrams at
the tree level.
The case of manifolds with non-trivial topology, in which zero modes may
 appear, is discussed in Section 3. It is shown that
zero modes generate large gauge transformations which leave invariant
the action and the equations of motion of the theories under
consideration. In this way, it becomes possible to treat zero modes as
gauge degrees of freedom and to gauge them away, as advocated in \cite{Pol}.
In Section 4 a set of BRST invariant  observables is derived and their
vacuum expectation values is computed.
Finally, the Conclusions and a possible physical application of the results
contained in this work
are presented in Section 5.
\section{Chern-Simons Field Theories with Non-Semisimple Group of Symmetry}
Let us consider a class of Chern-Simons field theories
with action:
\beq{S=\int_M\Omega_{iI}\epsilon^{\mu\nu\rho}(B_\mu^I\partial_\nu A^i_\rho+
\frac \lambda 6f_{jk}^IA_\mu^iA_\nu^jA_\rho^k)}{actionbase}
where $i,I=1,\ldots,N$ and $\Omega_{Ii}$ denotes a non-degenerate bilinear
form. 
 Summation over repeated
 indices is
everywhere understood. The theory is defined on a three dimensional
manifold $M$ without boundary and equipped with an Euclidean metric.
For simplicity, we assume for the moment that all the de Rham
cohomology groups $H^n(M)$ are trivial, so that the problem
of harmonic zero modes does not appear. We will discuss zero modes
in the next Section.

The action \ceq{actionbase} is invariant
under the following gauge transformations:
\begin{eqnarray}
A^i_\mu&\rightarrow& A^i_\mu+\partial_\mu\eta^i\label{gautraone}\\
B^I_\mu&\rightarrow&
B^I_\mu+\partial_\mu\theta^I-\lambda {f^I_{ij}}\left(
\frac{\eta^i\partial_\mu\eta^j}2+\eta^iA_\mu^j\right)\label{gautratwo}
\end{eqnarray}
for arbitrary functions $\eta^i$ and $\theta^I$.
The above transformations correspond to a non-semisimple group of
symmetry. The related generators
$X_i$ and $H_I$ satisfy the following
 non-semisimple Lie algebra:
\beq{[H_I,H_J]=[H_I,X_j]=0\qquad\qquad[X_i,X_j]=f^I_{ij}H_I}{algebra}
with structure constants $f^I_{ij}$. This Lie algebra consists in an
abelian Lie algebra $\mathfrak{g}$ 
with a central extension 
by an abelian group
$\mathfrak{h}$. The generators of $\mathfrak{g}$  and $\mathfrak{h}$
are
the $X_i$ and $H_I$'s respectively. Let us denote with the symbols
$\mathfrak{G}$  and $\mathfrak{H}$ respectively
the associated abelian Lie groups.
The matrix $\Omega_{Ii}$
appearing
in \ceq{actionbase} is the
generalization of
the Killing form to the case of non-semisimple groups.
Theories such as those discussed here have been already 
proposed in \cite{wit,tse}. A thorough discussion of their
renormalization
has been provided in \cite{grassi}. Other  applications of
non-semisimple Lie algebras can be found in \cite{nssalg}.

To eliminate the gauge freedom of the action \ceq{actionbase},
we  introduce the covariant gauge conditions:
\beq{\partial^\mu A_\mu^i=\partial^\mu B_\mu^I=0}{gaucond}
The theory can now be quantized using the procedure of BRST
quantization.
The BRST transformations associated to the gauge
transformations (\ref{gautraone}--\ref{gautratwo}) are given by:
\begin{eqnarray}
\delta A_\mu^i&=&\partial_\mu c^i\label{brst1}\\
\delta B_\mu^I&=&\partial_\mu \xi^I
+\lambda f^I_{ij}A^i_\mu c^j
\label{brst2}\\
\delta \xi^I&=&\frac\lambda2f^I_{ij}c^ic^j\label{brst3}\\
\delta c^i&=&0\label{brst4}\\
\delta\bar c_i&=&ia_i\qquad\qquad \delta a_i=0\label{brst5}\\
\delta\bar \xi_I&=&ib_I\qquad\qquad \delta b_I=0\label{brst6}
\end{eqnarray}
where $c^i,\xi^I$ and $\bar c_i,\bar \xi_I$ are anti-commuting ghosts,
while $a_i,b_I$ are scalar fields. It is possible to verify that the
transformations (\ref{brst1}--\ref{brst6}) are nilpotent,
i. e. $\delta^2=0$. 

At this point, one can write the expression of the gauge fixed BRST
invariant action:
\beq{S_{BRST}=S+S_{gf}+S_{FP}}{quaaction}
where $S$ is given by Eq.~\ceq{actionbase}, while the gauge fixing and
Fadeev-Popov terms are respectively:
\beq{S_{gf}=i\int_M d^3x\left[a_i\partial^\mu A^i_\mu+b_I\partial^\mu
B^I_\mu \right] }{gaufix}
\beq{S_{FP}=\int_M d^3x\left[\partial_\mu \bar c_i\partial^\mu
c^i+\partial^\mu \bar \xi_I\left(\partial_\mu \xi^I+\lambda
f^I_{ij}A_\mu^ic^j\right )\right]}{fadpop}
The combination $S_{gf}+S_{FP}$ amounts to a BRST exact
variation as expected:
\beq{S_{gf}+S_{FP}=-\delta\int_M\left[(\bar c_i\partial^\mu
A^i_\mu+
\bar\xi_I\partial^\mu B^I_\mu\right]}{totbrstvar}
Thus, the gauge fixing and Faddeev-Popov terms do not
spoil the topological properties of the original action
\ceq{actionbase}.

Let us note that
the fields
$B^I_\mu$ and $\xi^I$ play in \ceq{quaaction}
the role of pure Lagrange multipliers, which
constrain the fields $A^i_\mu$ and $\bar \xi_I$ in such a way
that all
 possible radiative
corrections vanish identically.
In particular, the interaction term in the
ghost
action \ceq{fadpop} disappears after an integration over the fields
$\xi^I$, which
 gives as a result the constraints
\beq{\partial_\mu\partial^\mu\bar\xi_I=0}{constrone}
Choosing suitable boundary conditions for which the $\bar\xi_I$'s
do not diverge at infinity, the  above equation is satisfied only for
constant 
 fields $\bar\xi_I$. The integration over the fields $B^I_\mu$ leads
instead to the flatness conditions:
\beq{\Omega_{Ii}\epsilon^{\mu\nu\rho}\partial_\nu A^i_\rho=0}{constrtwo}
The above equations determine the transverse components of the fields
$A^i_\rho$, while the longitudinal components are fixed by the gauge
fixing \ceq{gaufix}.

\begin{figure}[h]
\vspace{1 truein}
\includegraphics{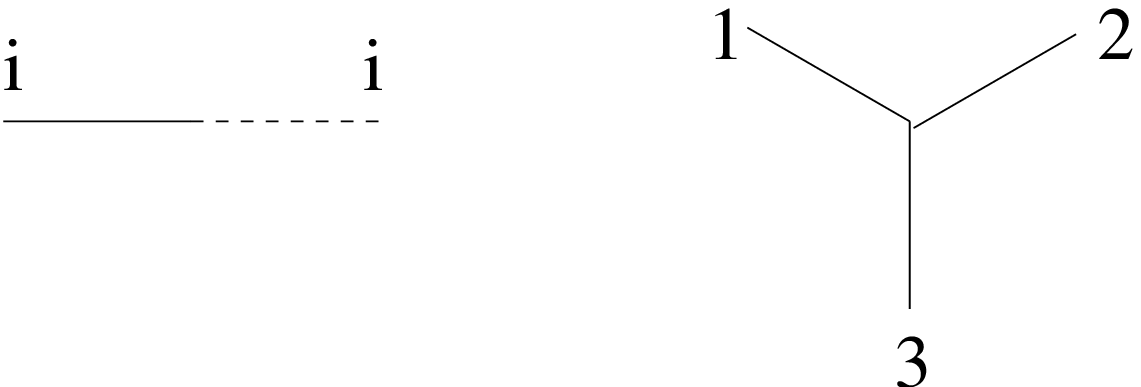}
\vspace{0.2in}
\label{feyrul}
\end{figure}

Since radiative corrections are absent, the theory \ceq{quaaction} is
purely classical. Thus, contrarily to what happens for instance in the 
case of Chern-Simons theories with gauge group $SU(N)$,
there is no rescaling of the
 Chern-Simons coupling 
constants, which here have been set equal to one.
Also at the classical
 level
several simplifications occurs and in practice the theory admits only the two
connected Feynman diagrams shown in Fig.~\ref{feyrul}.
These diagrams correspond to the field
propagators and to the three-vertex associated to the
cubic interaction term present in
Eq.~\ceq{quaaction}. Higher order tree diagrams, which  could
in principle be constructed by contracting together
the legs of many three-vertices, 
are actually ruled out due to the off-diagonal structure of the
propagators, which forbids any self-interaction among
the fields $A^i_\mu$. 

Due to the fact that the theory is purely classical, its partition function
can be exactly derived  once the classical solutions of the field equations are
known.
With a simple integration it is possible to eliminate
the fields $B^I_\mu$ and the ghosts. As an upshot, one obtains the constraints
 (\ref{constrone}--\ref{constrtwo}). These constraints and the gauge fixing
relations are enough
 to determine uniquely the remaining fields. If the theory is defined 
for instance on a
 manifold with flat metric, the solutions of the constraints
 \ceq{constrtwo} are simply  $A^i_\mu=0$, so that the
 partition function is the trivial one.
\section{The Zero Mode Problem}
In this Section we
consider the case
in which the fields admit non-trivial
classical configurations, the so-called harmonic zero modes.
We can ignore possible zero modes in the sectors of the ghost fields
and of the Lagrange multipliers $a_i,b_I$, because these zero modes are
not used in the gauge fixing procedure. We are thus left only with the
zero modes of the fields $A_\mu^i,B_\mu^I$, which we denote with the
symbols $\alpha_\mu^i$ and $\beta_\mu^I$ respectively.
From the action \ceq{actionbase} one finds the relevant equations of
motion which define $\alpha_\mu^i$ and $\beta_\mu^I$:
\begin{eqnarray}
\epsilon^{\mu\nu\rho}\partial_\nu A_\rho^i&=&0\label{equone}\\
\epsilon^{\mu\nu\rho}\partial_\nu B_\rho^I&+&\frac
\lambda2\epsilon^{\mu\nu\rho}f^I_{jk} A_\nu^jA_\rho^k=0\label{equtwo}
\end{eqnarray}
The general solution of Eq.~\ceq{equone} is
\beq{A_\mu^i=\alpha_\mu^i+\partial_\mu\eta^i}{solone}
where $\partial_\mu\eta^i$ is an exact differential, while $\alpha_\mu^i$ is a
 non-trivial abelian flat connection corresponding to the
abelian subgroup of the underlying gauge group.
Let us consider now Eq.~\ceq{equtwo}. This relation can be rewritten
as follows:
\beq{\epsilon^{\mu\nu\rho}\partial_\nu B^I_\rho=J^{I\mu}}{auxone}
with the current $J^{I\mu}$ given by:
\beq{J^{I\mu}=-\frac
\lambda2\epsilon^{\mu\nu\rho}f^I_{jk}A_\nu^jA_\rho^k}{currjimu} 
Eq.~\ceq{equone} implies that $J^{I\mu}$ is purely transverse,
i. e. $\partial_\mu J^{I\mu}=0$.
Using this fact, it is possible to show that
 Eq.~\ceq{equtwo} is solved by:
\beq{B_\mu^I=\int_M d^3y
G_{\mu\nu}(x,y)J^{I\nu}(y)+\beta^I_\mu+\partial_\mu\eta^I
}{soltwo}
where $G_{\mu\nu}(x,y)$ is the propagator of the theory in the Lorentz
gauge \ceq{gaucond} and
$\beta^I_\mu$ is a non-trivial flat 
connection satisfying the flatness condition
$\epsilon^{\mu\nu\rho}\partial_\nu \beta_\rho^I=0$. We
remember that
the fields $B_\mu^I$ are pure Lagrange
multipliers imposing the constraints \ceq{equone} on the fields
$A^i_\mu$. It is easy to check that the presence or not of the term
$\beta_\rho^I$ does not affect these constraints nor the other
equations of motion, so that one can put $
\beta^I_\mu=0$ without any loss of generality.

A possible strategy to treat harmonic zero modes is to consider them
as gauge degrees of freedom and to gauge them away using BRST techniques.
This approach has been proposed by Polyakov in \cite{Pol} and further
developed in \cite{proharzer}. An application to the 
$BF-$systems can be found in \cite{blauthom}.
In order to check if it is possible to translate the zero mode problem
in a gauge fixing problem
also in the present case, the crucial
point is to verify the invariance of the theory \ceq{actionbase} under
large gauge transformations.
As a matter of fact,
the $\alpha_\mu^i$'s  generate large gauge
transformations, consisting in multivalued mapping of the manifold $M$
onto the elements of the abelian group which corresponds to the Lie algebra
$\mathfrak{G}$ defined after Eq.~\ceq{algebra}.

To express the large gauge transformations acting on the fields
in a closed form, it is
convenient to introduce
potentials
$\Lambda^i$ such that:
\beq{\partial_\mu\Lambda^i=\alpha_\mu^i}{potdef}
These potentials, which will be in general multivalued on the
manifold
$M$, are the analogous of the functions $\eta^i$
appearing in 
the gauge transformations (\ref{gautraone}--\ref{gautratwo}). 
In terms of the $\Lambda^i$'s, the large gauge transformations induced
by the harmonic modes $\alpha_\mu^i$ are given by:
\begin{eqnarray}
A^i_\mu&\rightarrow&
A^i_\mu+\partial_\mu\Lambda^i\label{largautraone}\\
B^I_\mu&\rightarrow&
B^I_\mu-\lambda {f^I_{ij}}\left(
\frac{\Lambda^i\partial_\mu\Lambda^j}2+\Lambda^iA_\mu^j\right)
\label{largautratwo}
\end{eqnarray}
Actually, the full transformations of the fields $B^I_\mu$ are not necessary
in order to discuss the gauge invariance
of the action $S$ and of the equations of motion. As a matter of fact, 
apart from a total derivative,
the action \ceq{actionbase} can
be rewritten as follows:
\beq{S=\int_M
d^3x\Omega_{Ii}\epsilon^{\mu\nu\rho}A^i_\mu\left[\partial_\nu
B^I_\rho+\frac\lambda 6 f^I_{jk}A_\nu^jA_\rho^k\right]}{newforact}
Thus, only the transformations of the
pseudo-tensors $\epsilon^{\mu\nu\rho}\partial_\nu
B^I_\rho$ are needed:
\begin{eqnarray}
\epsilon^{\mu\nu\rho}\partial_\nu B^I_\rho&\rightarrow&
\epsilon^{\mu\nu\rho}\partial_\nu
B^I_\rho\nonumber\\
&-&\frac\lambda2\epsilon^{\mu\nu\rho}
 {f^I_{jk}}\left(
\partial_\nu\Lambda^j\partial_\rho\Lambda^k+2\Lambda^j\partial_\nu A^k_\rho
+2\partial_\nu\Lambda^jA_\rho^k\right)\label{auxlargautratwo}
\end{eqnarray}
At this point we are ready to perform a large gauge transformation of
the kind (\ref{largautraone}--\ref{auxlargautratwo})
in the action $S$.
Since the right hand side of Eq.~\ceq{auxlargautratwo}
has an explicit dependence on the potentials
$\Lambda^i$, the gauge transformed action will contain multivalued
contributions. However, it is easy to prove that
all these contributions vanish identically
due to the
following identities, valid up to total derivative terms which
are irrelevant 
on a manifold without boundary:
\begin{eqnarray}\int_Md^3x\Omega_{Ii}
\epsilon^{\mu\nu\rho}f^I_{jk}A^i_\mu\partial_\rho
A_\nu^j\Lambda^k &=&-\frac 12\int_M
d^3x\Omega_{Ii}\epsilon^{\mu\nu\rho}f^I_{jk}A^i_\mu
A^j_\nu\alpha_\rho^k\label{ideone}\\
\int_M d^3x\Omega_{Ii}\epsilon^{\mu\nu\rho}f^I_{jk}\alpha_\mu^i\partial_\rho
A^j_\nu\Lambda^k&=&-\int_M
d^3x\Omega_{Ii}\epsilon^{\mu\nu\rho}f^I_{jk}\alpha_\mu^i
A^j_\nu\alpha^k_\rho
\label{idetwo}
\end{eqnarray}
Another important identity, which follows from the fact that
the $\alpha_\mu^i$'s satisfy the classical equations of motion
\ceq{equone}, is:
\beq{\int_M d^3x
\Omega_{Ii}\epsilon^{\mu\nu\rho}f^I_{jk}\alpha_\mu^i\alpha_\nu^j
\alpha_\rho^k= \int_M d^3x
 \Omega_{Ii}\epsilon^{\mu\nu\rho}f^I_{jk}\partial_\mu\Lambda^i
\partial_\nu\Lambda^j\partial_\rho\Lambda^k=0}{idethr}
With the help of the relations (\ref{ideone}--\ref{idethr}) it is
possible to verify the invariance of the action $S$ under large gauge
transformations as desired. 

One can also check that the gauge transformations
(\ref{largautraone}) and \ref{auxlargautratwo}) preserve the form of
the equations of motion (\ref{equone}--\ref{equtwo}).
After a gauge transformation, in fact, one obtains the following
result:
\begin{eqnarray}
\epsilon^{\mu\nu\rho}\partial_\nu A_\rho^i&=&0\label{gtequone}\\
\epsilon^{\mu\nu\rho}\partial_\nu B_\rho^I&+&\frac
\lambda2\epsilon^{\mu\nu\rho}f^I_{jk}\left(
A_\nu^jA_\rho^k-2\partial_\nu A_\rho^k \Lambda^j\right)
=0\label{gtequtwo}
\end{eqnarray}
The spurious term proportional to $\Lambda^j$ in Eq.~\ceq{gtequtwo}
vanishes identically due to Eq.~\ceq{gtequone}.

\section{Observables and Wilson Loop-like amplitudes}
Good observables of a topological field theory should be BRST
 invariant and
lead
 to
vacuum expectation values which are metric
 independent.

To derive a set of  observables for the theory under consideration, we 
first observe that the following quantity is invariant
under
 the BRST
 transformations (\ref{brst1}--\ref{brst6}):
\begin{eqnarray}
T^I(\Gamma)&=&\oint_\Gamma dx^\mu
B^I_\mu+\frac\lambda{4\pi}f^I_{ij}\oint_\Gamma dx^\mu A^i_\mu \int d^3
y\frac 1{|x-y|}\partial_y^\rho A^j_\rho(y)\nonumber\\
&{\!\!\!\!\!\!\!\! \!\!\!\!\!\!\!\! \!\!\!\!\!\!\!\!
\!\!\!\!\!\!\!\!\!\!\!\!\!\!\!\!\!\!
\!\!\!\!\!\!
+}&\!\!\!\!\!\!\!\! \!\!\!\!\!\!\!\! \!\!\!\!\!\!\!\! \!\!\!\! 
 \frac\lambda{2(4\pi)^2}f^I_{ij}\oint_\Gamma dx^\mu\int
d^3y\left(\partial_\mu^x \frac 1{|x-y|}\right)\partial_y^\rho A^i_\rho
(y) \int d^3z\frac 1{|x-z|}\partial_z^\sigma
A_\sigma^j(z)\label{obstwo} 
\end{eqnarray}
In the above formula as well as in the rest of this paper, it has been
assumed
for simplicity that the
manifold $M$ coincides with the three dimensional euclidean space {\bf R}$^3$. 
The form of $T^I(\Gamma)$ has been obtained starting from the line
integral $\oint_\Gamma dx^\mu B^I_\mu$   and adding suitable terms in
order to make it gauge and BRST invariant.
At this point, for $M$ loops $\Gamma^a$, $a=1,\ldots,M$,
 it is possible to write down analogs of the holonomic
connections as follows:
\beq{W(\bar C)=e^{iC_{aI}T^I(\Gamma^a)}}{anaholcon}
where $\bar C$ is a matrix having as elements constant
parameters $C_{aI}$.

We note that $W(\bar C)$ is of the form:
\beq{W(\bar C)=\exp\left[iC^{aI}\oint_{\Gamma_a} dx^\mu
B^I_\mu+\int d^3x \chi_i(x)\partial^\mu A^i_\mu(x)\right]}{forcom}
All contributions coming from
the various line integrals which are present in the right hand side of
Eq.~\ceq{obstwo} are now contained in the scalars $\chi_i(x)$. 
As a consequence of Eq.~\ceq{forcom}, even if $W(\bar C)$ is manifestly metric
dependent,
a metric variation of this observable can always be
 compensated by a
shift of the Lagrange multipliers $a_i$ which impose the gauge
condition in the action \ceq{quaaction}. Thus, the vacuum expectation
values of the operators $W(\bar C)$ lead to metric independent
amplitudes as required.

When computing the vacuum expectation value $\langle W(\bar C)\rangle$
of the operator \ceq{obstwo}, several simplifications occur. One reason
is that the fields $A^i_\mu$ are unaffected by the cubic interactions
present in the action \ceq{quaaction}, because they can be contracted
only with the fields $B^I_\mu$ due to the off-diagonal form of the
kinetic terms. On the second hand, as already mentioned,
the fields $B^I_\mu$ behave as
abelian fields and act as Lagrange multipliers, which
impose the conditions:
\beq{\Omega_{Ii}\epsilon^{\mu\nu\rho}\partial_\nu
A_\rho^i(x)+\sum_{a=1}^M C_{Ia}\oint_{\Gamma^a}dy^\mu\delta(x-y)=0
}{const}
The solution of the above equation is given by:
\beq{A^{(cl)i}_\mu(x)=(\Omega^{-1})^{iI}C_{Ia}{\cal B}_\mu^a(x)
}{clasol}
with
\beq{{\cal B}_\mu^a(x)= -\frac{\epsilon_{\mu\sigma\tau}}{4\pi}
\oint_{\Gamma^a}dx^\sigma\partial^\tau\frac 1{|x-y|}}{magfie}
Let us note that $A^{(cl)i}_\mu(x)$ is a purely transverse vector
field because the longitudinal components have been fixed to zero by
the gauge condition \ceq{gaucond}. At this point, it is possible to
evaluate the explicit expression of $\langle W(\bar C)\rangle$ using
the saddle point evaluation method. After some calculations one finds:
\beq{\langle W(\bar C)\rangle=\exp\left[
\frac{i\lambda}6\int d^3x l_{abc}\epsilon^{\mu\nu\rho}{\cal B}^a_\mu(x)
{\cal B}^b_\nu(x){\cal B}^c_\rho(x)
\right]}{finres}
where
\beq{l_{abc}=((\Omega^{-1})^{jJ}(\Omega^{-1})^{kK}
C_{Ia}C_{Jb}C_{Kc}f_{jk}^I}{labc}
From Eq.~\ceq{finres} it turns out that
 the vacuum expectation values of the operators $W(\bar
C)$ 
 deliver essentially a single topological
invariant which is given by:
\beq{{\cal H}=\int d^3x l_{abc}\epsilon^{\mu\nu\rho}{\cal B}^a_\mu(x)
{\cal B}^b_\nu(x){\cal B}^c_\rho(x)}{topinv}
Another topological invariant, namely the Gauss linking number, can be
 obtained by considering also amplitudes containing Abelian
holonomic connections 
of the $A^i_\mu$ fields. Abelian connections are sufficient to grant BRST
 invariance in this case due to the simplicity of the
BRST transformations \ceq{brst1} of the fields
$A^i_\mu$.
\section{Conclusions}
In this work we have investigated a class of topological field theories
having the property that its perturbative
series contains only the finite set of Feynman diagrams given in
Fig.~\ref{feyrul}.
These theories are exactly solvable and, besides the Gauss
link invariant which is typical of
Abelian C-S field theories,  produce
 the topological
invariant ${\cal H}$ of Eq.~\ceq{topinv}. 
The fact that a non-Abelian Chern-Simons field theory can sustain Abelian
 observables like the $W(\bar C)$ of Eq.~\ceq{obstwo} is  related to
the presence of a non-trivial Abelian ideal in the gauge group of symmetry.

It is interesting to consider the above results from the point of view of
possible applications to the statistical mechanics of random walks. Let
$\Gamma^1,\ldots,\Gamma^M$ be a set of $M$ closed random walks, interacting
together via the topological potential \ceq{topinv}. The partition function
of this system
coincides with the sum over all possible configurations of the trajectories
$\Gamma^1,\ldots,\Gamma^M$:
\beq{Z=\int{\cal D}\br_1(s_1)\ldots{\cal D}\br_M(s_M)
 \exp\left[\sum_{a=1}^M\oint_{\Gamma^a}{\dot {\br}}_a^2+
{\cal H}(\br_1,\ldots,\br_M)\right]}{rwparfun}
where $\br_a(s_a)$, $a=1,\ldots,M$ are  curves described in the space by
 the paths $\Gamma^a$ and
parametrized by means of their arc-lengths $s_a$.
At this point the relation \ceq{finres} may be interpreted as the analog of
an Hubbard-Stratonovich transformation \cite{ferannphys},
which decouples the trajectories
$\Gamma^a$ in the partition function $Z$. Such transformation
 simplifies the task of
performing the path integration in Eq.~\ceq{rwparfun}, which is otherwise
very complicated due to the presence of the topological term
${\cal H}(\br_1,\ldots,\br_M)$. Other applications can be in the 
study of the fractional quantum Hall effect \cite{qhe,mdwp}.

To conclude, one should mention that the idea
of constructing a topological field theory  with a finite number
of  Feynman diagrams has already been realized following a different route.
This is the so-called Rozansky-Witten topological sigma model, which
delivers topological invariants of the Milnor type of its
hyper-K\"ahler target space \cite{rozwit}.
Interesting new developments in this
direction have been presented in \cite{blatho,broda}.
\section{Acknowledgments}
I wish to thank heartily the referee of this paper, G. Thompson, whose
contribution has led to considerable
improvements with respect to the original version. In particular, I am
grateful to G. Thompson for pointing out that the theory in
Eq.~\ceq{actionbase} is related to a Chern--Simons field
theory with non-semisimple group of gauge symmetry and for suggesting
Eqs.~(\ref{gautraone}--\ref{gautratwo}), which provide the full set of
gauge transformations of the theory. 
I am also grateful to G. Thompson for having written a very
enlightening referee report on this paper and for his interest in my
work. 
Finally, I wish to thank him for drawing my attention to the stimulating
problem of the treatment of harmonic zero modes.

\newpage
\section{Figure Captions}
\mbox{}
\\[1cm]
\begin{description}
\item[Fig. 1] Feynman rules corresponding to the action \ceq{quaaction}.
Dashed lines propagate $B_\mu^I$ fields while solid lines
are associated to $A^i_\mu$ fields.
\end{description}
\newpage
\section{Figures}
\mbox{}
\\[1cm]
\mbox{}\\[1cm]
\centerline{\Large Fig. 1}
\end{document}